# Cluster Based Hierarchical Routing Protocol For Wireless Sensor Network

Md. Golam Rashed[1], M. Hasnat Kabir[2], Muhammad Sajjadur Rahim[3], Shaikh Enayet Ullah[4]

[1] Department of Electronic and Telecommunication Engineering (ETE),
University of Development Alternative (UODA), Bangladesh .

[2,4] Department of Information and Communication Engineering
University of Rajshahi, Rajshahi-6205, Bangladesh.

[3] School of Information Engineering, University of Padova, Italy.

**Abstract:** *The efficient use of energy source in a sensor node is most desirable criteria for prolong the life time of wireless sensor network. In this paper, we propose a two layer hierarchical routing protocol called Cluster Based Hierarchical Routing Protocol (CBHRP). We introduce a new concept called head-set, consists of one active cluster head and some other associate cluster heads within a cluster. The head-set members are responsible for control and management of the network. Results show that this protocol reduces energy consumption quite significantly and prolongs the life time of sensor network as compared to LEACH.*

## 1. Introduction

Wireless Sensor Networks (WSN) consists with huge tiny sensors that are arranged in spatially distributed terrain. Due to a large number of applications such as security, agriculture, automation and monitoring, WSN has been identified as one of the pioneer technique in 21$^{st}$ century. The deployment and maintenance of the small sensor nodes should be straightforward and scalable when they are useless. But the small sensor nodes are usually inaccessible to the user, and thus replacement of the energy source is not feasible. There are some technical constrains also in microsensor node such as low energy in battery and processing speed etc. This limitation should be overcome as much as possible to use the node in wireless sensor network. The source of energy in sensor is a main constrain among them [1]. It can be minimized by increasing the density of energy in conventional energy sources [2],[3]. One of the elegant ways to prolong the lifetime of the network is to be reduced energy consumption in sensor nodes. This idea has been applied many research [4],[5] to increase the life time of the network. Cluster based Low-Energy Adaptive Clustering Hierarchy (LEACH) is an attractive routing protocol that disperses the energy load among the sensor nodes [6],[7],[8]. The improvements of LEACH have been presented elsewhere [9],[10],[11]. Proposed protocol is another approach to improve of LEACH protocol by reducing energy consumption and prolong the life time of the network.

## 2. The Proposed Approach (CBHRP)

In this paper an optimum energy efficient cluster based hierarchical routing protocol for wireless sensor network is proposed, which is a two layer protocol where a number of cluster cover the whole region. Proposed protocol introduces a concept of head-set instead of a cluster head. At one time, only one member of head-set is active and the remaining are in sleep mode. Several states of a node are found in this protocol such as- *candidate state, non-candidate state, active state, associate state, and passive associate state*. This protocol divides the network into a few real clusters including an active cluster head and some associate cluster heads. For a given number of data collecting nodes, the head-set members are systematically adjusted to reduce the energy consumption, which increases the network life.

## 3. Architecture of CBHRP

In the proposed model, the number of clusters k and nodes n are pre-determined for the wireless sensor network. Iteration consists of two stages: an election phase, and a data transfer phase. At the beginning of election phase, a set of cluster heads are chosen on random basis. These cluster heads send a short range advertisement broadcast message. The sensor nodes receive the advertisements and choose their cluster heads based on the signal strength of the advertisement messages. Each sensor node sends an acknowledgment message to its cluster head. Moreover, in each iteration, the cluster heads choose a set of associate heads based on the signal strength of the acknowledgments. A head-set consists of a cluster head and the associates. The head-set member is responsible to send messages to the base station. Each data transfer phase consists of several epochs. Each member of head-set becomes a cluster head once during an epoch. A round consists of several iterations. In one round, each sensor node becomes a member of head-set for one time. All the head-set members share the same time slot to transmit their frames.

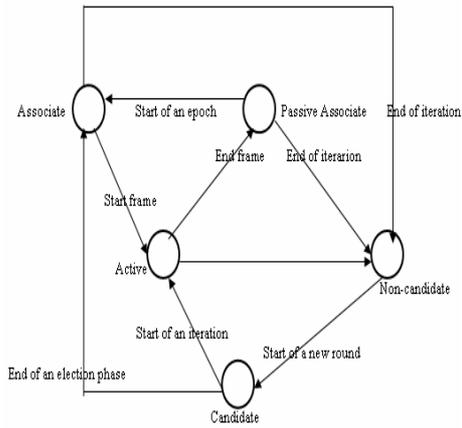

**Figure 1**. Different states of a sensor node in CBHRP protocol.

Different states of a sensor node in wireless sensor network are shown in Figure 1. The damaged or malfunctioning sensor states are not considered. Each sensor node joins the network as a candidate. At the start of each iteration, a fixed number of sensor nodes are chosen as cluster heads; these chosen cluster heads acquire the active state. By the end of election phase, a few nodes are selected as members of the head-sets within a cluster; these nodes acquire associate state where one of them is in active state and the remaining is in associate state. In an epoch of a data transfer stage, the active sensor node transmits a frame to the base station and goes to the passive associate state. At that time the next associate member acquires the active state. Therefore, during an epoch, the head-set members are distributed as follows: one member is in active state, a few members are in associate state, and remaining are in passive associate state. At the time of last frame transmission of an epoch, one member is active and the remaining are passive associates; there is no member in an associate state. For the start of next epoch, one head-set member acquires active state and the remaining are associate. By completing iteration, all the head-set members acquire the non-candidate state. The members in non-candidate state are not eligible to become a member of a head-set. For starting a new round, all non-candidate sensor nodes acquire candidate state.

## 4. Result and Discussion:

### 4.1 Energy consumption

The energy consumption for specific number of frames and different head set-sizes with respect to the variation of cluster number, network diameter size, and distance of the base station from the network are examined.

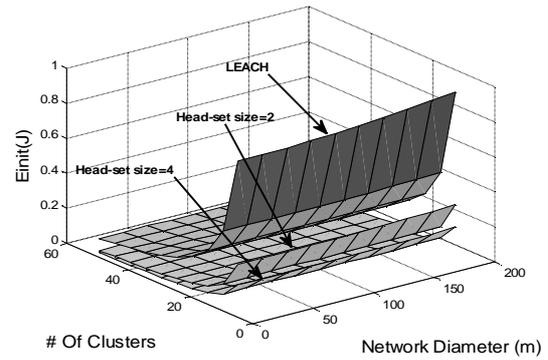

**Figure 2.** Consumed energy per round with respect to number of cluster and distance of base station.

Figure 2. shows that the variation of energy consumption per node with respect to the number of clusters and network diameter. Figure shows that energy consumption is reduced when the number of cluster are increased. It also shows that the optimum range of cluster between 20 and 60. It is interesting to point out that the number of head-sets reduces the consumption of energy as compared to LEACH because it reduces the election process.

Figure 3. illustrates the difference of the energy consumed per round with respect to the head set size and network diameter. It is seen that energy consumption is reduced when the head set size is increased. It can be concluded from above two figures that the head-set members play significant role to reduce the energy consumption than that of LEACH.

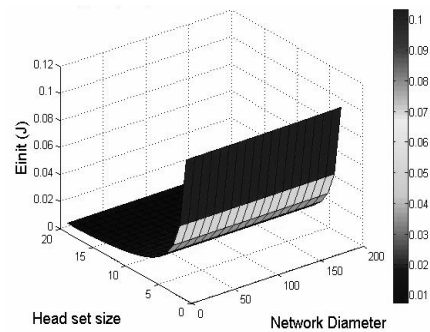

**Figure 3.** Consumed energy consumed per round wrt network diameter and head-set size when base station is at fixed location.

### 4.2 Iteration time and frames

The estimated time for one iteration with respect to the network diameter considering the percentage of head-set size is shown in figure 4. The iteration time is proportional to the initial energy $E_{init}$ and the network diameter found in this figure. The network will be alive for a longest period of time with initial energy when the head-set size is 50% of the cluster size. However, it is more or less with respect to the head-set size.

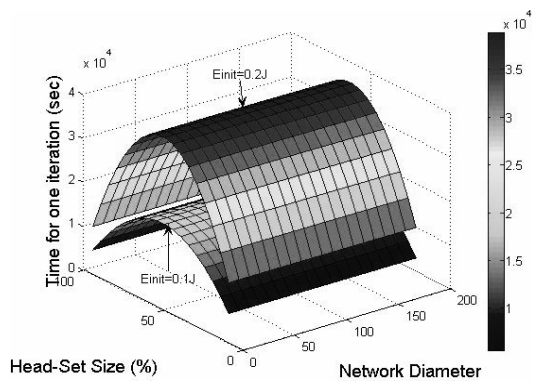

**Figure 4.** Time for iteration with respect to network diameter and head-set size.

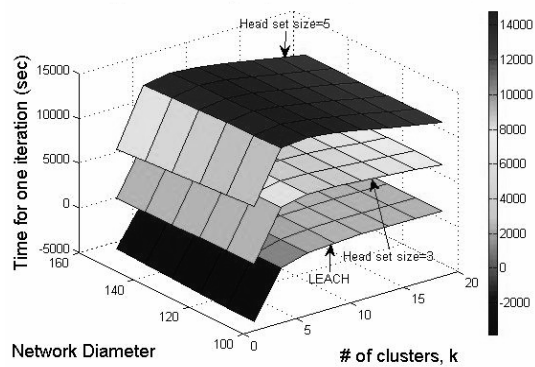

**Figure 5.** Time for iteration with respect to number of clusters and network diameter.

Figure 5 shows a graph that illustrates the estimated time for one iteration with respect to the number of clusters and network diameter. The number of cluster does not affect on iteration time for a particular network diameter. However, the head-set size plays a significant role to increase the iteration time. In contrast to LEACH, iteration time is increased by several times for a given number of head-set.

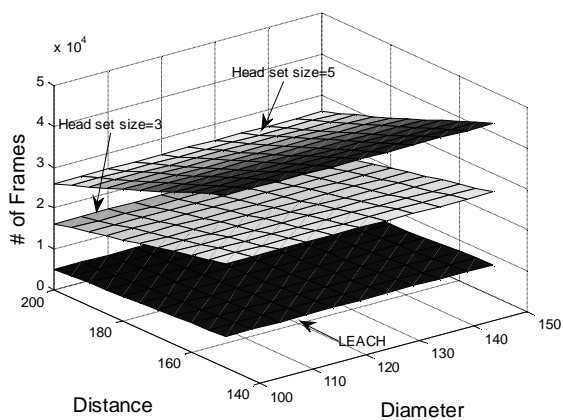

**Figure 6.** Number of frames transmission per iteration with respect to distance of BS and network diameter.

Data transmission in a wireless network is one of the most important issues. It is a measuring tool of a network suitability and perfectness. Our results show that the proposed method is able to transmit higher data frame in contrast to LECAH for the same dimension and distance of BS. While, head-set size is responsible to increase the data frame transmission. On the other hand, it is clearly found that the transmission rate slightly decreases with respect to the distance of BS. Figure 6 shows the number of frames transmitted with respect to network diameter and distance from BS.

## 5. Conclusion

An effective routing protocol is designed and tested it performances to overcome some present limitation of WSN. Introducing head-set concepts instead of only one cluster head within a cluster, our results show the better performance than that of LEACH in context of energy consumption, frame transmition and the life time of the sensor network.

## Authors Profile

**Md. Golam Rashed** received the B.Sc. and M.Sc. degree in Information and Communication Engineering from University of Rajshahi, Bangladesh in 2008 and 2009, respectively. He is currently working as a Lecturer in the department of Electronic and Telecommunication Engineering(ETE), University of Development Alternative (UODA), Bangladesh .

**M. Hasnat Kabir** received the B.Sc. and M.Sc. degree in Applied Physics and Electronics from University of Rajshahi, Bangladesh in 1995 and 1996, respectively. He has completed his PhD from Kochi University of Technology, Japan in 2007. Now he is working as an Assistant Professor in the dept. of Information and Communication Engineering, University of Rajshahi, Bangladesh.

**M. Sajjadur Rahim** received his B.Sc. (Engg.) degree in Electrical and Electronic Engineering from Bangladesh University of Engineering and Technology, Bangladesh and Master of Engineering degree in Information Science and Systems Engineering from Ritsumeikan University, Japan in 1999 and 2006, respectively. He is working as an Assistant Professor in the dept. of Information and Communication Engineering, University of Rajshahi, Bangladesh. At present, he is pursuing his PhD degree in the School of Information Engineering at the University of Padova, Italy on study leave.

**Shaikh Enayet Ullah** received the B.Sc. and M.Sc. degree in Applied Physics and Electronics from University of Rajshahi, Bangladesh. He has completed his PhD from Jahangirnagor University, Bangladesh. He is currently working as a Professor and Chairman of the dept. of Information and Communication Engineering, University of Rajshahi, Bangladesh.